# Superconductivity induced by Mg deficiency in non-centrosymmetric phosphide Mg$_2$Rh$_3$P

Akira Iyo,[1*] Izumi Hase,[1] Hiroshi Fujihisa,[1] Yoshito Gotoh,[1] Nao Takeshita,[1] Shigeyuki Ishida,[1] Hiroki Ninomiya,[1] Yoshiyuki Yoshida,[1] Hiroshi Eisaki[1], Kenji Kawashima,[1,2]

[1]National Institute of Advanced Industrial Science and Technology (AIST), 1-1-1 Umezono, Tsukuba, Ibaraki 305-8568, Japan

[2]IMRA Material R&D Co., Ltd., 2-1 Asahi-machi, Kariya, Aichi 448-0032, Japan

**ABSTRACT:** The search for non-centrosymmetric superconductors that may exhibit unusual physical properties and unconventional superconductivity has yielded the synthesis of a non-centrosymmetric phosphide Mg$_2$Rh$_3$P with an Al$_2$Mo$_3$C-type structure. Although stoichiometric Mg$_2$Rh$_3$P does not exhibit superconductivity at temperatures above 2 K, we found that an Mg deficiency of approximately 5 at.% in the Mg$_2$Rh$_3$P induced superconductivity at 3.9 K. Physical properties such as the lattice parameter $a$ = 7.0881 Å, Sommerfeld constant $\gamma_n$ = 5.36 mJ mol$^{-1}$ K$^{-2}$, specific heat jump $\Delta C_{el}/\gamma_n T_c$ = 0.72, electron–phonon coupling constant $\lambda_{e-p}$ = 0.58, upper critical field $H_{c2}(0)$ = 24.3 kOe, and pressure effect d$T_c$/d$P$ = −0.34 K/GPa were measured for the superconducting Mg$_{2-\delta}$Rh$_3$P ($\delta \sim 0.1$). Band-structure calculations indicate that exotic fermions, which are not present in high-energy physics, exist in Mg$_2$Rh$_3$P. Since Mg, Rh, and P are the first elements used at each crystal site of Al$_2$Mo$_3$C-type compounds, the discovery of Mg$_2$Rh$_3$P may guide the search for new related materials.

## I. INTRODUCTION

Non-centrosymmetric superconductors (NCS) whose crystal structures lack inversion symmetry have attracted significant attention because of their potential as a platform for novel physical phenomena, as well as the possibility for unconventional superconductivity. [1,2] Besides the exploration of intriguing physical properties, it is also important to expand the variety of NCS compounds. Compounds of the form $A_2M_3X$, with an Al$_2$Mo$_3$C-type structure ($P4_132$, space group no. 213), do not have inversion symmetry. Figure 1 shows the schematic crystal structure of the $A_2M_3X$ compound type. Its structure consists of a three-dimensional network of corners sharing $M_6X$ octahedrons. The $M_6X$ octahedrons are significantly distorted because two $A$ atoms are located in the space surrounded by the octahedrons. When one $A$ atom occupies the space, the structure becomes a so-called "anti-perovskite $AXM_3$" that typically has the $P$m-3m space group.

Compounds (a)–(g) in Table 1 are $Al_2Mo_3C$-type superconductors identified before this study. These superconductors are attractive in terms of both theoretical physics and application-based perspectives. The superconducting state of $Li_2Pt_3B$ has been shown to include a spin-triplet component by nuclear magnetic resonance (NMR) spectroscopy [3], while $Li_2Pd_3B$ appears to be spin-singlet. $Al_2Mo_3C$ is a strong-coupled superconductor that deviates from Bardeen–Cooper–Schrieffer (BCS)-type behavior [4], with a relatively high critical temperature ($T_c$ = 9.3 K) and high upper critical field $H_{c2}(0)$ (= 150 kOe) [5]. Moreover, the $A_2M_3X$ crystal structure is attracting interest, as it has been shown that a new type of quasiparticle referred to as an exotic fermion exists in $Li_2Pd_3B$ [6]. This new type of quasiparticle may provide a novel platform for topological quantum devices.

As shown in Table 1(a)–(g), the $M$ site in $A_2M_3X$ superconductors is occupied by a 4$d$–5$d$ transition metal (= Nb, Mo, Pd, Re, or Pt) from various element groups. In addition, the $A$ (= Li, Al, Cr, Rh, Ag) and $X$ (= B, C, N, S) sites accept various elements from a wide range of element groups. This indicates that the $A_2M_3X$ ternary system is flexible in terms of the combination of the three constituent elements, and still presents possibilities for exploring new superconductors.

We tested several possible combinations of $A$, $M$, and $X$ from a wide range of element groups, and consequently succeeded in synthesizing the new $Al_2Mo_3C$-type phosphide $Mg_2Rh_3P$. Though stoichiometric $Mg_2Rh_3P$ did not show superconductivity, we found that superconductivity was induced in $Mg_2Rh_3P$ by Mg deficiency, as listed in Table 1(h). In this paper, we first describe the synthesis conditions of non-superconducting stoichiometric samples (NS-$Mg_2Rh_3P$) and superconducting Mg-deficient samples (SC-$Mg_{2-\delta}Rh_3P$). We then experimentally evaluate the refined crystal structures and basic physical properties of the samples, such as magnetization, resistivity, specific heat, and pressure effects. Finally, we address the existence of the above-mentioned exotic fermions in $Mg_2Rh_3P$ using theoretical calculations.

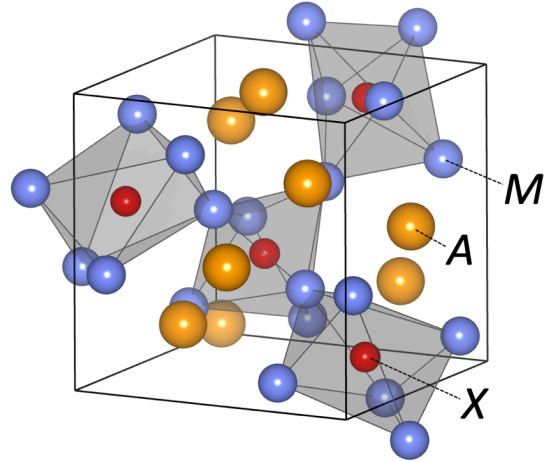

FIG. 1. Schematic of the $A_2M_3X$ ($Al_2Mo_3C$-type) crystal structure, produced using the VESTA software [7]. Thick solid lines indicate the unit cell.

Table I. Superconductors with the $Al_2Mo_3C$-type structure reported in the literature, and the $Mg_{2-\delta}Rh_3P$ superconductor synthesized in this study.

|     | Compounds | $T_c$ (K) | $a$ (Å) | $H_{c2}(0)$ (kOe) | Ref. |
|-----|-----------|-----------|---------|-------------------|------|
| (a) | $Al_2Mo_3C$ | 9.3 | 6.87 | 150 | 8,4,5 |
| (b) | $Al_2Nb_3N$ | 1.3 | 7.03 | - | 9 |
| (c) | $Ag_2Pd_3S$ | 2.25 | 7.23 | 6.3 | 10,11 |
| (d) | $Li_2Pd_3B$ | 8 | 6.75 | 50 | 12,13 |
| (e) | $Li_2Pt_3B$ | 2.8 | 6.76 | 20 | 14,13 |
| (f) | $Cr_2Re_3B$ | 4.8 | 6.61 | 100 | 15 |
| (g) | $Rh_2Mo_3N$ | 4.3 | 6.81 | 74.1 | 16 |
| (h) | $Mg_{2-\delta}Rh_3P$ | 3.9 | 7.09 | 24.3 | * |

* This study

## II. MATERIAL SYNTHESIS

Control of Mg (vapor) concentration during the sample synthesis was necessary to control the superconductivity and quality of the $Mg_2Rh_3P$ samples. We propose the following method for sample synthesis.

An Rh powder was ground with a P chunk using a mortar in a glove box. The powder, with a nominal composition of $Rh_3P$, was pressed into pellet form (~0.2 g) and sealed in an evacuated quartz tube (8 mm in inner diameter and 80 mm in length) with a certain molar ratio of Mg chips (~1 mm in size) as described below. The temperature of the quartz tube was raised from room temperature to 875−925 °C for 2 h and maintained for 8−16 h. During the heat treatment, Mg vapor was generated inside the quartz tube and reacted with both the pellet and the quartz tube. The Mg vapor eventually disappeared when the reactions between the Mg and the pellet and quartz tube were completed.

As discussed in more detail below, the superconductivity of $Mg_2Rh_3P$ was induced by Mg deficiency. Both the Mg deficiency and sample quality depended on the quantity of the Mg chips, heating temperature and time, and also possibly the inner dimensions of the quartz tube. If the quantity of Mg chips was large enough, e.g., with a molar ratio of Mg to $Rh_3P$ of approximately 8–10 (0.115–0.143 g Mg chips for the 0.2 g $Rh_3P$ pellet), a non-superconducting sample was obtained because the sample was exposed to Mg vapor for the entire heat-treatment period.

The molar ratio of Mg to $Rh_3P$, heating temperature, and treatment time of 5 (0.072 g Mg chips for the 0.2 g $Rh_3P$ pellet), 925 °C, and 16 h, respectively, provided the best conditions for obtaining a superconducting sample. Impurity phases formed by the decomposition of $Mg_2Rh_3P$ were then observed at the surface of the superconducting pellet, while significant decomposition was not observed in the case of the non-superconducting sample. It appears that the Mg deficiency was introduced from the surface of the pellet after the Mg vapor pressure was decreased. Thus, the presence of a decomposed surface is an indicator of a superconducting pellet. These results suggest that the decomposition of $Mg_2Rh_3P$ most likely occurs when the Mg deficiency exceeds a certain upper limit.

## III. EXPERIMENTS AND METHODS

Powder X-ray diffraction (XRD) patterns were obtained at room temperature using a diffractometer (Rigaku, Ultima IV) with $CuK\alpha$ radiation. Crystal structures of the samples were refined via Rietveld analysis using BIOVIA Materials Studio Reflex software (version 2018 R2) [17]. Magnetization ($M$) measurements were performed under magnetic fields ($H$) using a magnetic-property measurement system (Quantum Design, MPMS-XL7). Electrical resistivity and specific heat ($C$) were measured by the four-probe method, using a physical-property measurement system (Quantum Design, PPMS). The compositions of the samples were analyzed by energy-dispersive X-ray (EDX) spectrometry (Oxford, SwiftED3000) using an electron microscope (Hitachi High-Technologies, TM3000). The $T_c$ was measured under pressures of up to 1.5 GPa using a piston-cylinder pressure cell (Electro Lab, MLPC-15). Daphne oil 7373 was used as a pressure-transmitting medium. The applied pressure was determined by a Pb manometer that was placed adjacent to the sample.

First-principles electronic band structure calculations were also performed using the full-potential linearized augmented plane-wave method and generalized gradient approximation for the exchange-correlation potential [18,19]. Spin-orbit interaction was included in the calculations. Details of our calculations are provided in a previous work [20].

## IV. RESULTS AND DISCUSSION

The experimental results were obtained for NS-$Mg_2Rh_3P$ and SC-$Mg_{2-\delta}Rh_3P$ synthesized with Mg:$Rh_3P$ molar ratios, heating temperatures, and heating times of 8, 875 °C, and 10 h and 5, 925 °C and 16 h, respectively. The decomposed surface of the SC-$Mg_{2-\delta}Rh_3P$ pellet was removed prior to the measurements. Both samples were silver in color and stable in ambient air.

### A. Composition

In order to elucidate the difference in superconductivity, the sample compositions were analyzed by EDX spectrometry. The composition ratio was determined by averaging the data from approximately thirty spots on the polished sample surfaces. In this manner, the obtained composition ratios for Mg:Rh:P were 1.90 (4):3.00 (4):1.04 (2) for SC-$Mg_{2-\delta}Rh_3P$, and 2.00 (4):3.00 (4):1.06 (1) for NS-$Mg_2Rh_3P$. SC-$Mg_{2-\delta}Rh_3P$ was found to have an Mg deficiency of approximately 5 at.% ($\delta \sim 0.1$), while NS-$Mg_2Rh_3P$ had a nearly stoichiometric composition ratio (Mg:Rh:P = 2:3:1). Considering the synthesis conditions of the samples, these results are in line with the expectations. It is likely that $Mg_2Rh_3P$ decomposes when the Mg deficiency exceeds ~5 at.% (upper limit for Mg deficiency).

### B. Crystal structure

The XRD patterns of NS-$Mg_2Rh_3P$ and SC-$Mg_{2-\delta}Rh_3P$ were similar, and could not be distinguished at a glance. Figure 2 shows the XRD pattern and Rietveld fitting for NS-$Mg_2Rh_3P$. The sample is nearly monophasic and its diffraction pattern is well fitted by assuming the $Al_2Mo_3C$-type structure. Small peaks from $Rh_2P$ are observed in the diffraction pattern. The results of Rietveld structure refinements for NS-$Mg_2Rh_3P$ and SC-$Mg_{2-\delta}Rh_3P$ are summarized in Table II. SC-$Mg_{2-\delta}Rh_3P$ has a slightly smaller lattice parameter than NS-$Mg_2Rh_3P$, which may arise from the Mg deficiency. Note that the Mg deficiency is too small to be detected by Rietveld analysis.

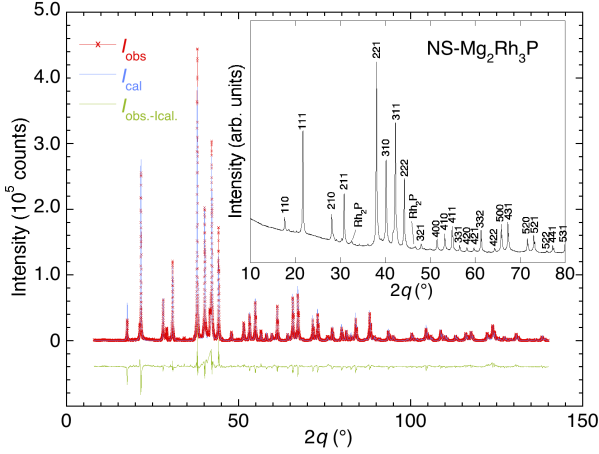

FIG. 2. Powder XRD pattern and Rietveld fitting for NS-$Mg_2Rh_3P$. $I_{obs.}$ and $I_{cal.}$ indicate the observed and calculated diffraction intensities, respectively. Inset: magnification diffraction angles $2\theta = 10-80°$, with a diffraction index for each peak.

Table II. Results of Rietveld structure refinements for NS-$Mg_2Rh_3P$ and SC-$Mg_{2-\delta}Rh_3P$ at room temperature.

|  | NS-$Mg_2Rh_3P$ | SC-$Mg_{2-\delta}Rh_3P$ |
|---|---|---|
| $a$ (Å) | 7.0900(1) | 7.0881(1) |
| $V$ (Å$^3$) | 356.35(2) | 356.11(2) |
| $x$ | 0.0397(2) | 0.0403(2) |
| $y$ | 0.1818(1) | 0.1823(1) |
| $l_{P-Rh}$ (Å) | 2.276 (1) | 2.274 (1) |
| $R_{wp}$ (%) | 13.10 | 14.00 |
| $R_e$ (%) | 9.05 | 11.03 |
| $S$ | 1.45 | 1.27 |

The space group is either $P4_132$ (Cubic, no. 213) or its enantiomorph, $P4_332$ (Cubic, no. 212). The occupancy is fixed to 1 at all atomic sites. The atomic coordinates of Mg, Rh, and P are $(x, x, x)$, $(1/8, y, y + 1/4)$, and $(3/8, 3/8, 3/8)$, respectively, while global isotropic temperature factors for all atoms are employed in both refinements. The mean-square displacements $U_{iso}$ (Å$^2$) are 0.024(1) and 0.025(1) for all atomic sites for NS-$Mg_2Rh_3P$ and SC-$Mg_{2-\delta}Rh_3P$, respectively.

## C. Superconductivity

Figure 3 shows the temperature ($T$) dependence of $4\pi M/H$ for NS-$Mg_2Rh_3P$ and SC-$Mg_{2-\delta}Rh_3P$. NS-$Mg_2Rh_3P$ shows a trace of superconductivity, which is not visible in Fig. 3. The small superconducting component is owing to the Mg deficiency in part of the polycrystalline $Mg_2Rh_3P$. However, a significant superconducting transition is observed for SC-$Mg_{2-\delta}Rh_3P$, with an onset $T_c$ of 3.94 K. The shielding volume fraction is calculated to be 98% at 2 K considering the demagnetizing effect. The volume fraction is large enough for the compound to be regarded as a bulk superconductor. The $T$ dependence of the magnetization under magnetic fields up to 10 kOe is shown in the inset of Fig. 3. $T_c$ is decreased with increasing applied magnetic field. The effect of Mg deficiency on superconductivity has previously been discussed by Hase et al. on the basis of electronic band structure calculations [20]. Their work indicated that the density of states (DOS) of NS-$Mg_2Rh_3P$ decreased at the Fermi level $E_F$, and that the value of the DOS was very small at $E_F$. Mg deficiency was shown to cause holes in the valence band and the consequent shift of $E_F$ drastically increased the DOS. Thus, the experimental result that only Mg-deficient $Mg_2Rh_3P$ shows superconductivity may be explained in terms of an increase in the DOS.

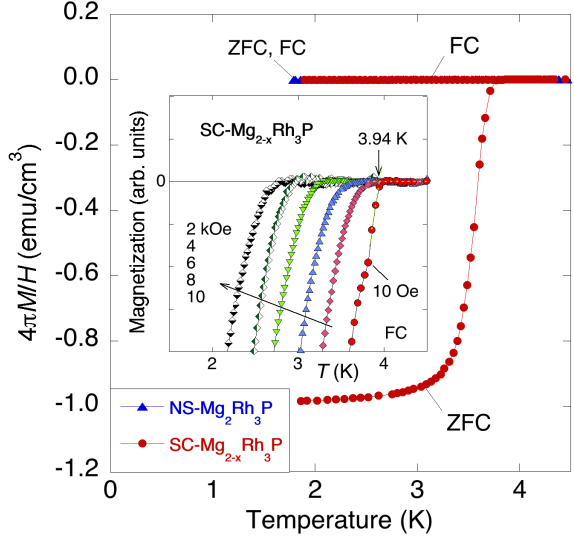

FIG. 3. $T$ dependence of $4\pi M/H$ for NS-Mg$_2$Rh$_3$P and SC-Mg$_{2-\delta}$Rh$_3$P. Measurements were performed with zero-field-cooling (ZFC) and field-cooling (FC) modes. Inset shows $T$ dependence of FC magnetization for SC-Mg$_{2-\delta}$Rh$_3$P under magnetic fields up to 10 kOe. Backgrounds were subtracted from the data.

Figure 4 shows the $T$ dependence of resistivity for NS- and SC-Mg$_{2-\delta}$Rh$_3$P. Both samples exhibit similar metallic $T$-dependent resistivity in the normal state. The residual resistance ratio $\rho(300\ \text{K})/\rho(\sim 0\ \text{K})$ of SC-Mg$_{2-\delta}$Rh$_3$P is smaller than that of NS-Mg$_2$Rh$_3$P, which can be explained by carrier scattering resulting from the Mg deficiency of SC-Mg$_{2-\delta}$Rh$_3$P. As shown in the inset of Fig. 4, only SC-Mg$_{2-\delta}$Rh$_3$P indicates a clear superconducting transition at the onset $T_c$ of 3.86 K, close to the value measured through magnetization tests. NS-Mg$_2$Rh$_3$P shows only a small decrease in resistivity below approximately 3.8 K, because of a low-content superconducting component detected by the magnetization measurement.

To approximate the Debye temperature $\Theta_D$, we fitted $\rho(T)$ above 50 K using the parallel resistance model $1/\rho(T) = 1/\rho_{BG}(T) + 1/\rho_{max}$ [21], where $\rho_{BG}(T)$ and $\rho_{max}$ are the Bloch–Grüneisen term and saturation resistivity, respectively. Fits employing this model are shown in Fig. 4 as yellow curves, revealing $\Theta_D$ = 335 K and 357 K for NS-Mg$_2$Rh$_3$P and SC-Mg$_{2-\delta}$Rh$_3$P, respectively. Note that the resistivity below 50 K cannot be fitted with this model.

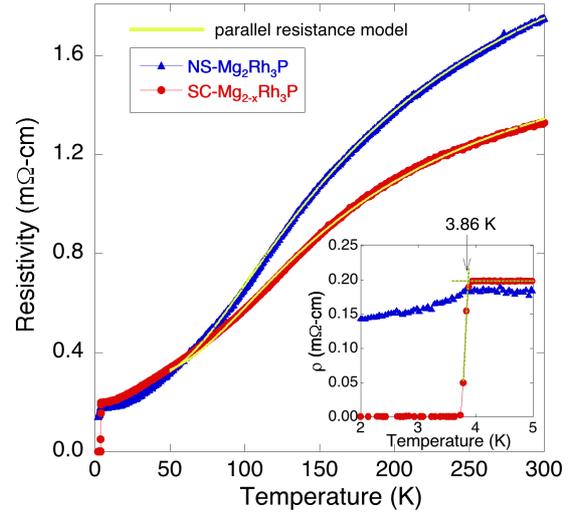

FIG. 4. $T$ dependence of resistivity for NS-Mg$_2$Rh$_3$P and SC-Mg$_{2-\delta}$Rh$_3$P. The inset shows an enlargement near the superconducting transition temperature. The yellow curves show fitting using the parallel resistance model.

The $T$ dependence of the resistivities under magnetic fields up to 14 kOe is shown in Fig. 5 for SC-Mg$_{2-\delta}$Rh$_3$P. $T_c$ decreases with an increasing applied magnetic field and a small magnetoresistance is observed. The $T$ dependence of the upper critical magnetic field $H_{c2}$ was obtained by defining the onset transitions of resistivity and magnetization, as shown in Figs. 5 and 3, respectively. The $H_{c2}$ values obtained from both measurements show good agreement.

The inset of Fig. 5 shows the reduced critical field $h^*$ ($= -H_{c2}/(dH_{c2}/dt)_{t=1}$) vs. $t$ ($= T/T_c$) [22,23]. The value of $h^*$ changes almost linearly, and no saturation is observed in the measured temperature range. SC-Mg$_{2-\delta}$Rh$_3$P is presumed to be a weak-coupling BCS-type superconductor with an experimentally evaluated small $\lambda_{\text{e-p}}$ (~ 0.58), as discussed in the next subsection. Werthamer–Helfand–Hohenberg (WHH) theory [22,23] was used to fit the data. The WHH fit is indicated by the dashed curve in the inset of the figure. $H_{c2}(0)$ is estimated to be 24.3 kOe using $H_{c2}(0) = 0.69T_c(dH_{c2}/dT)_{Tc}$ (dirty limit). This $H_{c2}$ (0) is considerably smaller than the Pauli limit $H_P$ ($= 1.86 \times T_c$) of ~70 kOe [24,25], suggesting that the upper critical field is limited by orbital pair breaking (Maki parameter $\alpha_M < 1/\sqrt{2}$ [26]) in SC-Mg$_{2-\delta}$Rh$_3$P. If we examine the behavior of $H_{c2}$ in more detail, $H_{c2}$ appears to deviate slightly upward from the WHH curve below $t \sim 0.8$. Such behavior is also observed in some NSCs [11,27,28]. To obtain a more detailed analysis, measurements of $H_{c2}$ at lower temperatures (< 2 K) are necessary.

The Ginzburg–Landau coherence length $\xi_0$ is calculated to be 116 Å employing $H_{c2}(0) = \Phi_0/2\pi\xi_0^2$, where $\Phi_0$ is the magnetic flux quantum.

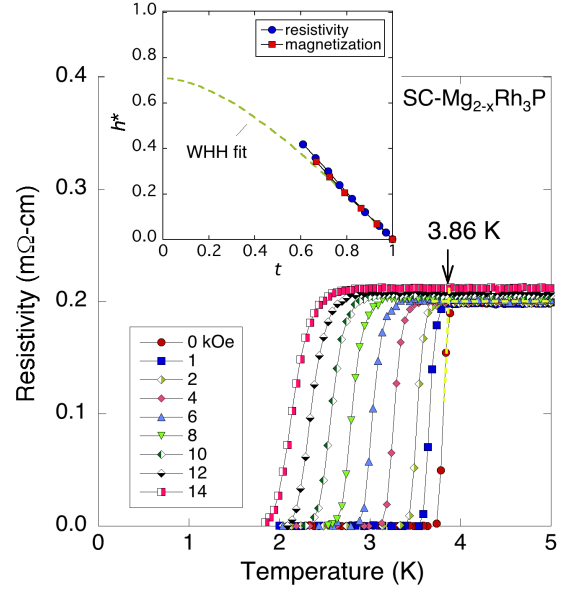

FIG. 5. $T$ dependence of the resistivity as a parameter of magnetic field strengths up to 14 kOe for SC-Mg$_{2-\delta}$Rh$_3$P. The $T_c$ was defined by the onset of transition. The inset shows the $t$ ($= T/T_c$) dependent reduced upper critical field $h^*$. The dashed curve indicates WHH fitting.

Figure 6 shows the magnetization ($M$–$H$) curves at various temperatures below $T_c$ for SC-Mg$_{2-\delta}$Rh$_3$P. The $M$–$H$ curves exhibit the typical behavior of type-II superconductors. To estimate the lower critical field $H_{c1}$, first, the $H_x$ is defined for each $M$–$H$ curve, as shown representatively for the $M$–$H$ curve at 2 K in Fig. 6. Then $H_{c1}$ is calculated by considering a demagnetization factor $N$ estimated to be 0.10 from the size of the measured sample, namely, $H_{c1} = H_p/(1-N)$. The inset of Fig. 6 depicts the thus-obtained $H_{c1}$ as a function of $(T/T_c)^2$. On the basis of GL theory $H_{c1}(T) = H_{c1}(0)[1-(T/T_c)^2]$, we derive $H_{c1}(0)$ to be 72.7 Oe. The London penetration depth of $\lambda_0 = 3010$ Å is calculated using an approximation of $H_{c1}(0) = \Phi_0/\pi\lambda_0^2$. The GL parameter of $\kappa_{GL} = \lambda_0/\xi_0$ is determined to be 26, which is an appropriate value ($> 1/\sqrt{2}$) for a type-II superconductor.

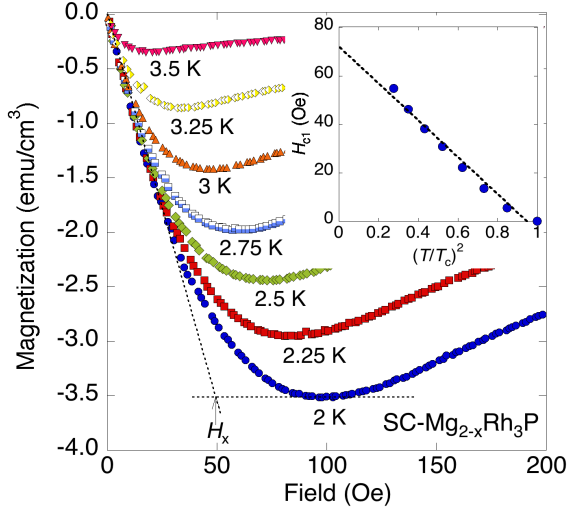

FIG. 6 Field dependence of magnetization at various temperatures below $T_c$ for SC-Mg$_{2-\delta}$Rh$_3$P. Inset shows $H_{c1}$ as a parameter of $(T/T_c)^2$. The dashed line in the inset is a linear fitting of the data.

### D. Specific heat

Figure 7 shows the $T^2$ dependence of $C/T$ for SC-Mg$_{2-\delta}$Rh$_3$P at $H = 0$. The data in the temperature range between 4.2−5 K is fitted with $C/T = \gamma_n + \beta T^2$, where $\gamma_n$ (the Sommerfeld constant) and $\beta$ are coefficients related to the electron and phonon contributions to the total specific heat ($C = C_{el} + C_{ph}$), respectively. The fitting yields values of $\gamma_n = 5.36$ mJ mol$^{-1}$ K$^{-2}$ and $\beta = 0.328$ mJ mol$^{-1}$ K$^{-4}$. The corresponding value of $\gamma_n$ for stoichiometric NS-Mg$_2$Rh$_3$P obtained by band-structure calculation was 1.41 mJ mol$^{-1}$ K$^{-2}$ [20]. This value is much smaller than the experimental result, which supports the concept that Mg deficiency causes an increase in the DOS of Mg$_2$Rh$_3$P. If we assume that Mg simply contributes electrons to the conduction band [20], we can derive the heat-capacity coefficient $\gamma_{band}$ for SC-Mg$_{2-\delta}$Rh$_3$P as 3.96 mJ mol$^{-1}$ K$^{-2}$, which is about 2.8 times larger than that of NS-Mg$_2$Rh$_3$P. We can estimate the mass enhancement factor as $\gamma_n/\gamma_{band} = 1.35$, which is quite reasonable for the mass enhancement by the electron–phonon interaction.

The Debye temperature $\Theta_D = 329$ K was derived using the formula $\beta = N(12/5)\pi^4 R \Theta_D^{-3}$, where R = 8.314 J/(mol K) and $N = 6$ ($N$ = the number of atoms in the unit cell). The $\Theta_D$ was close to the value obtained by the resistivity fitting. According to the McMillan equation for electron–phonon mediated superconductors [29], the electron–phonon coupling constant $\lambda_{e-p}$ can be determined by $\lambda_{e-p} = (\mu^* \ln(1.45 T_c/\Theta_D) - 1.04)(1 - 0.62\mu^*)/(1.04 + \ln(1.45 T_c/\Theta_D))$ where $\mu^*$ is a Coulomb pseudopotential parameter. Using $T_c = 3.86$ K, $\Theta_D = 329$ K, and the standard $\mu^*$ value (= 0.13), we obtained $\lambda_{e-p} = 0.58$, indicating that SC-Mg$_{2-\delta}$Rh$_3$P is a superconductor in the weak-coupling regime.

The inset in Fig. 7 shows the temperature dependence of $C_{el}/\gamma_n T$. The electronic specific heat $C_{el}$ is obtained by subtracting the phonon part ($C_{ph} = \beta T^3$) from the total $C$. A clear jump of $C_{el}/\gamma_n T$ is observed at 3.8 K (the midpoint temperature of the jump), which is in good agreement with the values measured by magnetization and resistivity, indicating the bulk nature of the superconductivity.

The dashed curve below $T_c$ in the inset in Fig. 7 shows the $T$-dependent $C_{el}/\gamma_n T$ calculated using weak-coupling BCS theory. The normalized specific heat jump $\Delta C_{el}/\gamma_n T_c$ is estimated to be 0.72, which is only approximately half of the BCS value (~1.43). Furthermore, the experimental $C_{el}/\gamma_n T$ values show more moderate temperature dependence than that predicted by the BCS theory. These findings suggest that the size of the superconducting gap $\Delta(0)$ of SC-$Mg_{2-\delta}Rh_3P$ is much smaller than the BCS value ($2\Delta(0)/k_B T_c = 3.53$). In order to understand the superconductivity in SC-$Mg_{2-\delta}Rh_3P$ in more detail, specific-heat measurements at lower temperatures (<2 K) are required.

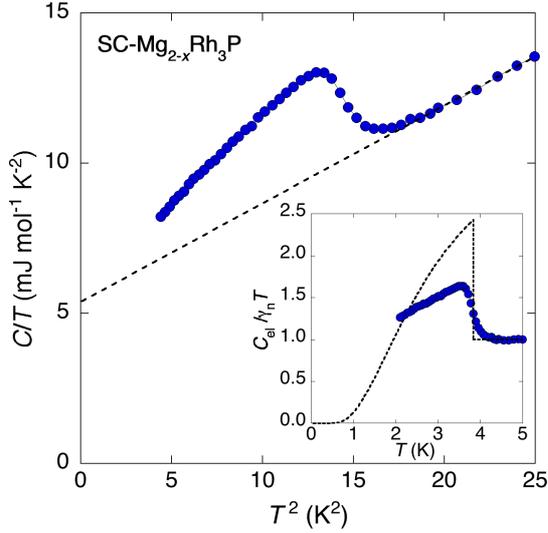

FIG. 7. $C/T$ vs $T^2$ of SC-$Mg_{2-\delta}Rh_3P$ below 5 K. The dotted line shows the fitting of data for $T = 4.2-5$ K with $C/T = \gamma_n + \beta T^2$ and its extrapolation to $T = 0$. Inset shows the temperature dependence of $C_{el}/\gamma_n T$. The $C_{el}/\gamma_n T$ below $T_c$ predicted by the BCS theory is shown as a dashed curve.

E. Pressure effects

The effects of pressure may help in enhancing $T_c$ of SC-$Mg_{2-\delta}Rh_3P$ through the substitution of the constituent elements of the compounds. We applied physical pressure to SC-$Mg_{2-\delta}Rh_3P$ to reveal the effect of lattice compression on $T_c$. Figure 8 shows the temperature dependence of normalized magnetization for SC-$Mg_{2-\delta}Rh_3P$ as a function of applied pressure. $T_c$ is observed to decrease with increasing pressure at a rate of approximately $dT_c/dP = -0.34$ K/GPa, as shown in the inset of Fig. 8. The decrease in $T_c$ can be simply explained by a decrease in the DOS at $E_F$ caused by the lattice size reduction under pressure. Conversely, $T_c$ may increase if a lattice parameter increases by the chemical substitution of an element such as Ca for Mg. The volume of SC-$Mg_{2-\delta}Rh_3P$ was expected to decrease typically ~1 % by applying a pressure of 1.5 GPa. However, the decrease in volume by the Mg deficiency was only 0.067% (see Table II). The effect of volume shrinkage due to Mg deficiency on DOS should be negligible. Thus, the emergence of superconductivity in $Mg_2Rh_3P$ (increase in DOS) is mainly attributed to the electron doping induced by the Mg deficiency.

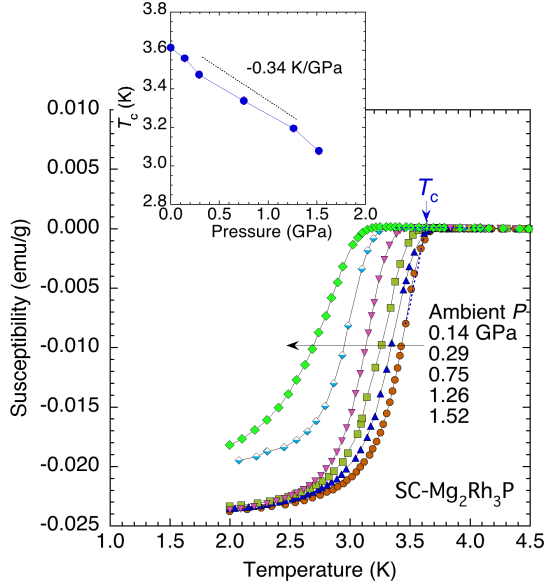

FIG. 8. Temperature dependence of magnetic susceptibility near $T_c$ for the SC-Mg$_{2-\delta}$Rh$_3$P sample, as a function of applied pressure ($P$). The inset shows the pressure dependence at the midpoint $T_c$.

### F. Exotic fermions in Mg$_2$Rh$_3$P

Dirac fermions are found in many compounds and are attracting increasing attention. Dirac fermions carry topological charge and are considered useful for quantum calculation. In general, electrons in a crystal have lower symmetry than in free space. Because of this lowered symmetry, various quasiparticles that are prohibited in free space can exist. Bradlyn et al. recently proposed the presence of extremely anomalous quasiparticles such as 3-fold, 6-fold, and 8-fold degenerated states in some crystal structures [6]. Figure 9 shows part of the band structure of NS-Mg$_2$Rh$_3$P near the Fermi level. The crystal structure of the NS-Mg$_2$Rh$_3$P belongs to space group #212 or #213, which can have 6-fold degenerated exotic fermions at the R($\pi/a$, $\pi/a$, $\pi/a$) point [6]. Along the R–X axis, all the bands are doubly degenerated because of the crystal symmetry. Moreover, each band along this line forms a "pair". In the limit of zero spin-orbit interaction, this splitting becomes zero. The size of the splitting is seen, for example, in the dashed square in Fig. 9. As shown, the magnitude of this splitting depends on the wave vector, but we can estimate the average splitting is similar to that in Li$_2$Pd$_3$B [6] and much smaller than that in Li$_2$Pt$_3$B [22, 30]. This result suggests that the parity mixing for the superconducting order parameter of SC-Mg$_{2-\delta}$Rh$_3$P may not be so large, likewise in Li$_2$Pd$_3$B. The state at the R point is 6-fold degenerated, and dispersion near this R point is approximately linear. This state may be observed by angle-resolved photoemission spectroscopy (ARPES) if a single crystal is obtained. Since this exotic fermion carries different topological charges than an ordinary Dirac fermion, if this state can be observed and controlled, it would be useful not only to explore novel physical properties, but also to create new types of topological quantum devices.

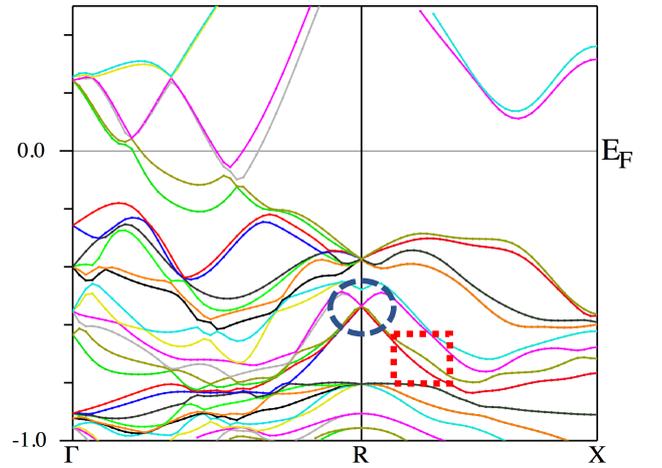

FIG. 9. Band structure near the Fermi energy ($E_F$) for NS-Mg$_2$Rh$_3$P. The unit of vertical axis is eV. The dashed circle shows the exotic fermion, which is the same type of quasiparticle as in Li$_2$Pd$_3$B [6]. The dashed square shows the band splitting arising from the lack of inversion symmetry and the spin-orbit interaction.

## V. CONCLUSION

We succeeded in synthesizing the new ternary phosphide Mg$_2$Rh$_3$P, with the Al$_2$Mo$_3$C-type structure, in both superconducting and non-superconducting states. We demonstrated that the superconductivity of Mg$_2$Rh$_3$P is induced by Mg deficiency, and that the superconductivity can be controlled by the synthesis conditions, which may be useful in device fabrication. We revealed the basic physical parameters of SC-Mg$_{2-\delta}$Rh$_3$P, as summarized in Table III, together with a determination of its electronic structure. There are issues regarding the behavior of $H_{c2}$ and the specific heat below 2 K that remain unanswered and require further study. Searching for new Mg$_2$Rh$_3$P-related compounds is also an interesting task left for a future study.

Table III. The physical parameters experimentally determined in this study for SC-Mg$_{2-\delta}$Rh$_3$P.

| Parameters | Values in SC-Mg$_{2-\delta}$Rh$_3$P |
| --- | --- |
| $T_c$ | 3.9 K |
| $H_{c1}$ | 73 Oe |
| $H_{c2}$ | 24.3 kOe |
| $\lambda_0$ | 3010 Å |
| $\xi_0$ | 116 Å |
| $\kappa_{GL}$ | 26 |
| $\gamma_n$ | 5.36 mJ mol$^{-1}$ K$^{-2}$ |
| $\beta$ | 0.328 mJ mol$^{-1}$ K$^{-4}$ |
| $\Theta_D$ | 329 K |
| $\lambda_{e-p}$ | 0.58 |
| $\Delta C_{el}/\gamma_n T_c$ | 0.72 |
| $dT_c/dP$ | –0.34 K/GPa |

**Acknowledgements**

This work was partially supported by JSPS KAKENHI Grant Number JP19K04481 and JP16H06439. We would like to thank Editage (www.editage.jp) for English language editing.